# Ultra-sensitive phase estimation with white light

Chuan-Feng Li,* Xiao-Ye Xu, Jian-Shun Tang, Jin-Shi Xu, and Guang-Can Guo

*Key Laboratory of Quantum Information, University of Science and Technology of China,
Chinese Academy of Sciences, Hefei, 230026, China*

(Dated: May 19, 2011)

An improvement of the scheme by Brunner and Simon [Phys. Rev. Lett. **105**, 010405 (2010)] is proposed to show that quantum weak measurements can provide a method to detect ultra-small longitudinal phase shifts even with white light. By performing an analysis in the frequency domain, we find that the amplification effect will work as long as the spectrum is large enough, irrespectively of the behavior in the time domain. As such, the previous scheme can be noteworthy simplified for experimental implementations.

High resolution phase estimation plays an important role in all fields of science where precise measurements of physical changes smaller than a particle's wave length are required. Usually, these phase estimations are realized with interferometers, which by exploiting the quantum coherence of light, have provided the basis for quantum metrology technologies [1] to measure small physical effects. Historically the first use was devised by Michelson in 1887 to measure the absolute motion of the Earth through the hypothetical ether [2]. A standard interferometric scheme exploits the fact that a half wave phase shift can change the pattern of interference fringes (interchanging dark and bright ports) and the intensity distribution of the light fringes indicates the amount of change in a physical variable that induces that phase shift. The resolution in a standard interferometer is dominated by the intrinsic quantum noise [3, 4]. To increase the resolution, researchers usually repeat the experiment many times. Recently, in exploring quantum features in a standard interferometer, some more sensitive phase estimation methods have been realized [5–8]. In a recent paper, based on quantum weak measurements and weak values [9], a promising alternative to standard interferometry in ultra-high resolution phase estimation has been proposed by Brunner and Simon (BS) [10].

One of the cornerstones of quantum mechanics is the famous measurement disturbance guaranteed by Heisenberg uncertainty principle [11], however, the concept of weak measurement has "opened the door for investigation of all manner of quantum phenomena previously deemed inaccessible" [12]. Weak measurement disturbs the measured system weakly and only extracts partial information about the system. However, combined with appropriate pre- and post-selections of the quantum state, strange weak values that lay outside the range of the observable's eigenvalues can be obtained. Theoretically, unorthodox predictions from weak measurements were initially controversial [13, 14]; experimentally, confirmation taken from various quantum systems, such as quan-tum optics [15, 16], solid state systems [17, 18] and quantum dots [19] have been outstanding. The weak perturbation in weak measurements of a quantum system can be very useful in the analysis of many interesting counterintuitive quantum phenomena, such as macrorealistic hidden variable theories [20–22], Hardy's paradox [23–25], apparent superluminal travel [26–28] and the three-box problem [29]. A more technical utilization bought about by weak measurements and weak values is the small signal amplification [12]. Compared to standard interferometry, weak value amplification not only provides the same precision as that when only pure real weak values are used, but can also exceed the precision when pure imaginary weak values are involved [27, 30]. The enhancement of a signal by using pure imaginary weak values has been used to detect the spin Hall effect of light [31] and tiny beam deflections [32–34]. The BS scheme proposed to measure small longitudinal phase shifts with a pulsed source is based on pure imaginary weak values. They point out that this method combined with frequency domain detection could in principle outperform standard interferometry by three orders of magnitude. Their theoretical analysis uses time domain analysis and a Fourier transform to show that a frequency shift is measurable given an imaginary weak value. In this report, we treat the weak measurement just as an interference and analyze it only in the frequency domain. As a result, we find that weak value amplification of small longitudinal phase shifts can be performed with classical thermal light with an ultra-broad bandwidth.

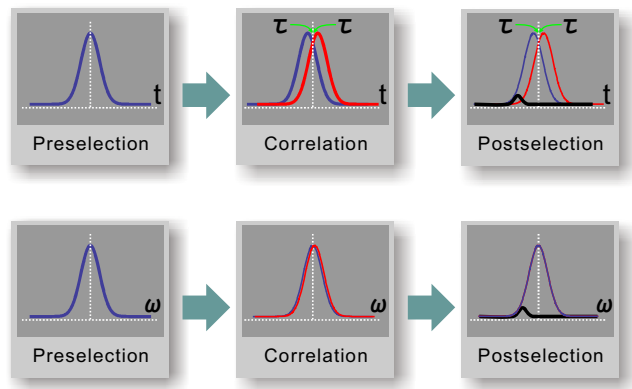

FIG. 1. (color online) Schematic of general weak measurements involving pure real weak values (the upper row) and imaginary weak values (the lower row).

* cfli@ustc.edu.cn

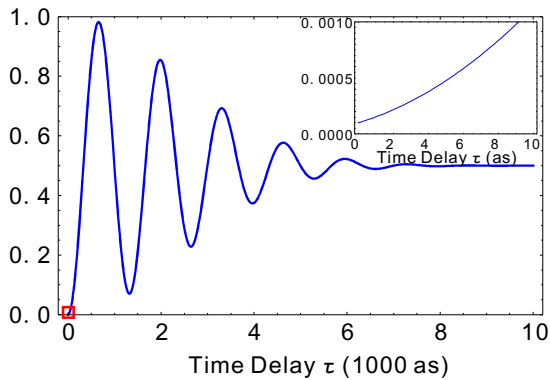

FIG. 2. (color online) Variation of the postselection probability $\mathcal{T}$ with time delay $\tau$, where $\epsilon$ is set at 0.01 radian and the spectral width is $100\,nm$. The red box and the inset identify the working range of weak measurements.

A standard quantum measurement procedure normally takes into account both the quantum system and the probe apparatus, and considers the wave packet collapse as a decoherence effect induced by the environment surrounding both [35]. A general weak measurement can be obtained through a change in a standard quantum measurement procedure by one of two distinct ways [36]: one is to keep the coupling strength the same as in a strong measurement but to change the initial state of the probe; the second is to keep the initial state of the probe but to reduce the coupling strength. These general weak measurement procedures are depicted in Figure 1. The quantum system considered is the polarization of a photon. Generally, the state after preselection can be written as $|\psi\rangle = \alpha|H\rangle + \beta|V\rangle$, where $|H\rangle$ ($|V\rangle$) represents the horizontal (vertical) polarization and $|\alpha|^2 + |\beta|^2 = 1$. The probe system can be another degree of freedom of the photon [10, 31–34] or an ancillary particle [16]. In the BS scheme, the probe is considered as the time of arrival of the single photon, expressed by $\int dt g(t)|t\rangle$, where $g(t)$ is the associated probe wave function and is assumed to be a Gaussian function $g(t) = (\pi\sigma^2)^{-1/4}\exp(-t^2/2\sigma^2)$ with $\sigma$ denoting the probe spread. A weak correlation is introduced through birefringence (which could be any other apparatus that has polarization dependent phase shift effects) and can lead to state changes of the probe in real space, that is, the time of arrival of $|H\rangle$ pulse and $|V\rangle$ pulse are shifted with $2\tau$ (the upper row in Fig. 1). After weak correlation, postselection on the system state

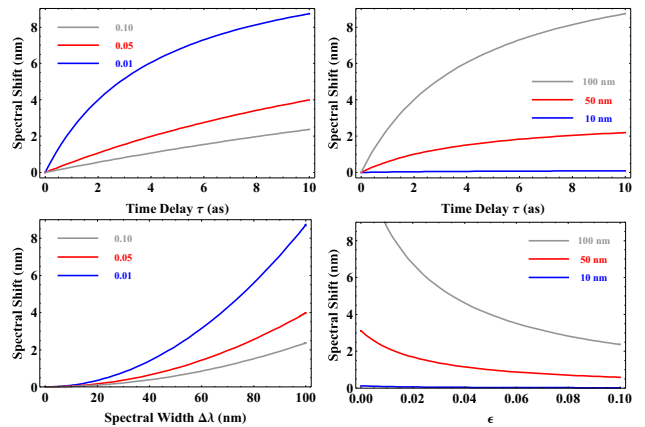

FIG. 3. (color online) Amplification effect involving imaginary weak values (see the text for detail).

can derive the so called weak value [9]

$$A_w = \frac{\langle\phi|A|\psi\rangle}{\langle\phi|\psi\rangle}, \qquad (1)$$

where $|\phi\rangle = \mu|H\rangle + \upsilon|V\rangle$ with $|\mu|^2 + |\upsilon|^2 = 1$ represents the postselection state, $A$ denotes the observable to be measured and $A|H\rangle = |H\rangle$, $A|V\rangle = -|V\rangle$. An appropriate selection of states $|\psi\rangle$ and $|\phi\rangle$ can yield weak values which lay outside the range of the observable's eigenvalues [9] or which are complex numbers [30]. The BS scheme takes $|\psi\rangle = \frac{1}{\sqrt{2}}(|H\rangle + i|V\rangle)$, $|\phi\rangle = \frac{1}{\sqrt{2}}(-ie^{i\epsilon}|H\rangle + e^{-i\epsilon}|V\rangle)$ and yield a pure imaginary weak value $A_w = i\cot\epsilon$ with postselection probability $\sin^2\epsilon$. The spectrum of the probe then equals $\sin^2(\omega\tau - \epsilon)|g(\omega)|^2$, where $g(\omega)$ is the Fourier transform of $g(t)$. The center of the probe spectrum will be shifted by $2\tau/\sigma^2\epsilon$. This shift in the light spectrum is very helpful in phase estimation when performing a spectral analysis. It is found that there is an amplification of the small longitudinal phase shifts. In their analysis, with a pulsed laser, the amplification is inversely proportional to the square of the temporal width of the pulses. However, through an analysis in only the frequency domain, we find that the amplification will work as long as the spectrum is large enough, irrespectively of the behavior in the time domain.

In our analysis for this report, the frequency of light is treated as the probe and the system state is set to the same as in the BS scheme. In the frequency wave function representation, the probe wave function is given by $f(\omega) = (\pi\delta^2)^{-1/4}\exp[-(\omega - \omega_0)^2/2\delta^2]$. The state of the system-probe combination after weak correlation can be written as

$$|\Psi\rangle = \int d\omega \frac{1}{\sqrt{2}} f(\omega)[e^{i\omega\tau}|H\rangle + ie^{-i\omega\tau}|V\rangle]|\omega\rangle. \qquad (2)$$

Although the weak measurement does not change the frequency distribution (the lower row in Fig. 1), there will be a spectrum shift when postselection of the system state has been performed. Postselection of the polarization state on basis $|\phi\rangle$ collapses the probe state to

$$|\mathcal{P}\rangle = \frac{1}{\sqrt{\mathcal{T}}} \int d\omega \frac{i}{2} f(\omega)[e^{i(\omega\tau-\epsilon)} + e^{-i(\omega\tau-\epsilon)}]|\omega\rangle, \quad (3)$$

with postselection probability

$$\mathcal{T} = 0.5\{1 - \exp(-\delta^2\tau^2)\cos[2(\omega_0\tau+\epsilon)]\}. \quad (4)$$

The imaginary weak value is given by $A_w = i\cot\epsilon$. The frequency distribution is given by

$$\mathscr{F}(\omega) = |\langle\mathcal{P}|\mathcal{P}\rangle|^2, \quad (5)$$

the center of which is

$$\omega_0^f = \frac{\int \omega \mathscr{F}(\omega)d\omega}{\int \mathscr{F}(\omega)d\omega} = \omega_0 + \Delta\omega, \quad (6)$$

where the frequency shift is

$$\Delta\omega = \frac{\tau\delta^2}{2\mathcal{T}}\exp(-\delta^2\tau^2)\sin[2(\omega_0\tau+\epsilon)]. \quad (7)$$

In the weak measurement limit, $\tau$ is extremely small. By taking the first order approximate, we then get $\mathcal{T} \to \sin^2\epsilon$ and $\Delta\omega \to \tau\delta^2\cot\epsilon$ (A factor 2 difference compared to the result of the BS scheme is because of a slightly different convention for the spread of the pointer). This frequency shift is measurable by currently available spectrometers taking into consideration alignment errors. Therefor the imaginary weak value amplification, it can be used to detect a small time delay $\tau$ which could in principle outperform standard interferometry by three orders of magnitude [10].

In Fig. 2, we compare the working ranges of a standard interferometer and weak measurements. The spectral width is $100\,nm$ for typical white light source. For such a large spectral width, postselection probabilities will quickly tend to 0.5 because of decoherence. For this reason, white light can not be used in a standard interferometer. However, a weak measurement works at the limit where $\tau$ is significantly smaller than the decoherence time characterized by $\delta$ (the red box and the inset in Fig. 2). For weak measurement, cooperation between preselection and postselection can discard most of the non-decohered part, which is dominant and acts as noise while the measured signal contained in quantum systems is ultra-small.

In Fig. 3, we depict the amplification arising through imaginary weak values. The top left corner displays the variation in spectral shifts with the time delay for different $\epsilon$ (Blue-solid, red-dotted and gray-dashed lines correspond to $\epsilon$ =0.01, 0.05 and 0.10, respectively. The spectral width is set to be $100\,nm$.) A smaller $\epsilon$, which indicates greater orthogonality, will be more suitable for measuring small phase shifts. However, it can not be set to zero as this corresponds to a zero count in the weak measurement limit. The variation of the spectral shifts with time delay for different spectral widths is depicted in the top right corner (Blue-solid, red-dotted and gray-dashed lines correspond to spectral width being $10\,nm$, $50\,nm$ and $100\,nm$, respectively. $\epsilon$ is set to 0.01.) A larger spectral width is significantly more suitable for amplification application. The lower two graphs present the variation in amplification factor with spectral width (lower left corner for different $\epsilon$) and $\epsilon$ (lower right corner for different spectral widths), where the time delay is fixed at $10\,as$.

In conclusion, an improvement of the BS scheme [10] has been proposed. While not involving real space, we derived a pure imaginary weak value and real frequency shift in the frequency domain and analyzed the amplification in measuring ultra-small longitudinal phase shifts. We noted that the original BS scheme can be significantly simplified and that this improved method can be realized in experiments by using just classical white light.

This work was supported by National Fundamental Research Program and the National Natural Science Foundation of China (Grant No.60921091, 10874162 and 11004185).

[1] V. Giovannetti, S. Lloyd, and L. Maccone, Phys. Rev. Lett. **96**, 010401 (2006).
[2] A. A. Michelson and E. W. Morley, Phil. Mag. **24**, 449 (1887).
[3] C. M. Caves, Phys. Rev. D **23**, 1693 (1981).
[4] B. Yurke, S. L. McCall, and J. R. Klauder, Phys. Rev. A **33**, 4033 (1986).
[5] K. Goda, O. Miyakawa, E. E. Mikhailov, S. Saraf, R. Adhikari, K. McKenzie, R. Ward, S. Vass, A. J. Weinstein, and N. Mavalvala, Nat. Phys. **4**, 472 (2008).
[6] R. Schnabel, N. Mavalvala, D. E. McClelland, and P. K. Lam, Nat. Commun. **1**, 121 (2010).
[7] T. Nagata, R. Okamoto, J. L. O'Brien, K. Sasaki, and S. Takeuchi, Science **316**, 726 (2007).
[8] B. L. Higgins, D. W. Berry, S. D. Bartlett, H. M. Wiseman, and G. J. Pryde, Nature **450**, 393 (2007).
[9] Y. Aharonov, D. Z. Albert, and L. Vaidman, Phys. Rev. Lett. **60**, 1351 (1988).
[10] N. Brunner and C. Simon, Phys. Rev. Lett. **105**, 010405 (2010).
[11] W. Heisenberg, Z. Phys. **43**, 172 (1927).
[12] A. M. Steinberg, Nature **463**, 890 (2010).
[13] I. M. Duck, P. M. Stevenson, and E. C. G. Sudarshan, Phys. Rev. D **40**, 2112 (1989).
[14] A. J. Leggett, Phys. Rev. Lett. **62**, 2325 (1989).
[15] N. W. M. Ritchie, J. G. Story, and R. G. Hulet, Phys. Rev. Lett. **66**, 1107 (1991).
[16] G. J. Pryde, J. L. O'Brien, A. G. White, T. C. Ralph, and H. M. Wiseman, Phys. Rev. Lett. **94**, 220405 (2005).
[17] A. N. Korotkov and A. N. Jordan, Phys. Rev. Lett. **97**, 166805 (2006).
[18] N. Katz, M. Neeley, M. Ansmann, R. C. Bialczak,

M. Hofheinz, E. Lucero, A. O'Connell, H. Wang, A. N. Cleland, J. M. Martinis, and A. N. Korotkov, Phys. Rev. Lett. **101**, 200401 (2008).
[19] A. Romito, Y. Gefen, and Y. M. Blanter, Phys. Rev. Lett. **100**, 056801 (2008).
[20] N. S. Williams and A. N. Jordan, Phys. Rev. Lett. **100**, 026804 (2008).
[21] A. Palacios-Laloy, F. Mallet, F. Nguyen, P. Bertet, D. Vion, D. Esteve, and A. N. Korotkov, Nat. Phys. **6**, 442 (2010).
[22] M. E. Goggin, M. P. Almeida, M. Barbieri, B. P. Lanyon, J. L. O'Brien, A. G. White, and G. J. Pryde, PNAS **108**, 1256 (2011).
[23] Y. Aharonov, A. Botero, S. Popescu, B. Reznik, and J. Tollaksen, Phys. Lett. A **301**, 130 (2002).
[24] J. S. Lundeen and A. M. Steinberg, Phys. Rev. Lett. **102**, 020404 (2009).
[25] K. Yokota, T. Yamamoto, M. Koashi, and N. Imoto, New J. Phys. **11**, 033011 (2009).
[26] D. R. Solli, C. F. McCormick, R. Y. Chiao, S. Popescu, and J. M. Hickmann, Phys. Rev. Lett. **92**, 043601 (2004).
[27] N. Brunner, V. Scarani, M. Wegmüller, M. Legré, and N. Gisin, Phys. Rev. Lett. **93**, 203902 (2004).
[28] Y. Aharonov and D. Rohrlich, *Quantum Paradoxes: Quantum Theory for the Perplexed* (Weinheim: Wiley-Vch, 2005).
[29] K. J. Resch, J. S. Lundeen, and A. M. Steinberg, Phys. Lett. A **324**, 125 (2004).
[30] R. Jozsa, Phys. Rev. A **76**, 044103 (2007).
[31] O. Hosten and P. Kwiat, Science **319**, 787 (2008).
[32] P. B. Dixon, D. J. Starling, A. N. Jordan, and J. C. Howell, Phys. Rev. Lett. **102**, 173601 (2009).
[33] D. J. Starling, P. B. Dixon, A. N. Jordan, and J. C. Howell, Phys. Rev. A **80**, 041803 (2009).
[34] D. J. Starling, P. B. Dixon, N. S. Williams, A. N. Jordan, and J. C. Howell, Phys. Rev. A **82**, 011802 (2010).
[35] W. H. Zurek, Rev. Mod. Phys. **75**, 715 (2003).
[36] Y. Kedem and L. Vaidman, Phys. Rev. Lett. **105**, 230401 (2010).